\documentclass[12pt]{article}
\usepackage{epsfig}
\newcommand{\bea}{\begin{eqnarray}}
\newcommand{\eea}{\end{eqnarray}}
\newcommand{\be}{\begin{equation}}
\newcommand{\ee}{\end{equation}}

\title{ Antishadowing in the rescaling model at $x \sim 0.1$
  \author{  A.V.~Kotikov$^{1,2}$, B.G.~Shaikhatdenov$^{2}$, Pengming Zhang$^{1,3}$ \\
    $^{1}$ Institute of Modern Physics, Lanzhou 730000, China\\
 $^{2}$ Joint Institute for Nuclear Research, Dubna 141980, Russia\\
 $^{3}$ University of Chinese Academy of Sciences, Yuquanlu 19A, Beijing 100049, China
 }}

\begin{document}
\maketitle
\abstract{
The antishadowing behavior of the sea and valence quark densities in nuclei is investigated in the framework of the
rescaling model at the values $x \sim 0.1$ of the Bjorken variable $x$.

{\it Keywords:} Deep inelastic scattering; parton densities; EMC effect.

\section{Introduction}

The analysis of deep-inelastic scattering (DIS) of leptons off nuclei
carried out in the valence quark dominance region and
first published by the European Muon Collaboration (EMC)~\cite{Aubert:1983xm}
 showed that there is a visible effect unexplainable by
the naive picture of a nucleus as being a bound system of quasi-free nucleons.
(for a review see, e.g.,~\cite{Arneodo:1992wf,Rith:2014tma}).

Nowadays there are two main approaches to studying this effect. In the first one, which is at present
more common, nuclear parton distribution functions (nPDFs) are extracted from the global fits (see a recent
review~\cite{Zurita:2018vrs} and references therein)
to nuclear data by using empirical parametrizations of their normalizations
and the numerical evaluation of
%solution to
Dokshitzer-Gribov-Lipatov-Altarelli-Parisi (DGLAP)
equations~\cite{DGLAP}.
The second strategy is based upon some models of nuclear PDFs (see
different models in, for example,~\cite{Kulagin:2004ie}-\cite{Close:1983tn} and a review~\cite{Kulagin:2016fzf}).

Here we will follow the rescaling model~\cite{Jaffe:1983zw,Close:1984zn}, which
%was very popular almost thirty years ago. The model
is based on suggestion~\cite{Close:1983tn} that the effective confinement size of gluons and quarks in the
nucleus is greater than in a free nucleon. In the framework of perturbative QCD it was found~\cite{Jaffe:1983zw,Close:1984zn,Close:1983tn}
that such a change in the confinement scale predicts that nPDFs and usual (nucleon) PDFs can be related by
simply rescaling their arguments.
%(see Eq.~(\ref{va.1a}) below).
Therefore, it can be said that the rescaling model demonstrates features that belong to both above approaches: in its framework there
are certain relations between usual and nuclear PDFs that result from shifting the values of kinematical variable $\mu^2$
and, at the same time, both densities obey DGLAP equations.

%At that time,
Originally, the rescaling model was established for the valence quark dominance region $0.2 \leq x \leq 0.8$.
Recently
%The aim of our paper is to extend
its applicability to the region of small $x$ values has been extended in Ref.~\cite{Kotikov:2017mhk},
where a certain shadowing effect has been found for the sea quark and gluon densities.

The aim of the present short paper is to extend our small $x$ results obtained in~\cite{Kotikov:2017mhk}
to the range $x \sim 0.1$ and to assess an antishadowing effect in that region.

\section{SF $F_2$}

The DIS structure function (SF) $F_2$ in the leading order (LO)
approximation, which is dealt with in the present paper,
has the following form
\begin{eqnarray}
  F_2(x,\mu^2) = e \,
  \bigl(
  f_q^S(x,\mu^2)+f_q^V(x,\mu^2) \bigr) + \Delta \, f_q^{NS}(x,\mu^2) ,~~~
\label{8a}
 \end{eqnarray}
where
$e=(\sum_1^f e_i^2)/f$ is an average of the squared quark charges, $\Delta$ is the difference
between charges of upper quarks and the average $e$, and $f_q^S(x,\mu^2)$, $f_q^V(x,\mu^2)$
 and $f_q^{NS}(x,\mu^2)$ are sea, valence and nonsinglet parts of quark parton density.

Below we will study only PDFs in the deuteron
%and the structure functions of the isoscalar nucleus
and, therefore, the contribution of the nonsinglet part $f_q^{NS}(x,\mu^2)$
can be omitted in the present analysis (see, for example, recent paper~\cite{Kotikov:2016ljf}
that discusses this possibility). So, below we will restrict ourselves to considering
sea and valence parts only.

Since we plan to extend the low $x$ PDF analysis obtained in~\cite{Kotikov:2017mhk} up to $x \sim 0.1$
the parton densities  $f_q^S(x,\mu^2)$ and $f_q^V(x,\mu^2)$ given above should be represented
in the following form \cite{Illarionov:2010gy}
%(R=S,V)
\be
f_q^R(x,\mu^2) = \tilde{f}_q^R(x,\mu^2) \, (1-x)^{\beta_R(s)}, ~~
s=\ln \left( \frac{a_s(\mu^2_0)}{a_s(\mu^2)} \right),~~
a_s(\mu^2) \equiv \frac{\alpha_s(\mu^2)}{4\pi} = \frac{1}{\beta_0\ln(\mu^2/\Lambda^2_{\rm LO})},
%~~~ (R=S,V)
\label{8aN}
 \ee
where hereafter $R=S,V$ and $\alpha_s(\mu^2)$ is the strong coupling constant.
In the above expressions the large $x$ asymptotics,
which starts to be important already at $x \sim 0.1$, is explicitly displayed.
We would like to note that here it is possible to use the following
%$\mu$-independent approximations:
\be
\beta_V(s) = \beta_V(0) + \frac{16 s}{3\beta_0},~~
\beta_V(0) \sim 3,~~  \beta_S(s) \sim \beta_V(s)+2 \, .
\label{bV}
 \ee
%which
The last two relations come from the quark counting rules \cite{Matveev:1973ra} and usually agrees with the
results of fits (see, for example, \cite{Kotikov:2016ljf,Krivokhizhin:2005pt}).

\subsection{Sea part}

At LO the small-$x$ asymptotic expressions for sea quark and gluon densities $f_a$
%(hereafter $\tilde{f}_q^S \equiv f_q$)
can be written as follows (the next-to-leading order (NLO) results can be found in Ref.
\cite{Q2evo}):
%--\cite{Cvetic1}):
\begin{eqnarray}
\tilde{f}_a(x,\mu^2) &=&
%~=~
f_a^{+}(x,\mu^2) + f_a^{-}(x,\mu^2),~~(\mbox{hereafter}~~~a=q,g) \nonumber \\
	f^{+}_g(x,\mu^2) &=& \biggl(A_g + \frac{4}{9} A_q \biggl)
		\tilde{I}_0(\sigma) \; e^{-\overline d_{+} s} + O(\rho), \nonumber \\
f^{+}_q(x,\mu^2) &=&
%&=&
\frac{f}{9} \biggl(A_g + \frac{4}{9} A_q \biggl) \rho \tilde{I}_1(\sigma)  \; e^{-\overline d_{+} s}
+ O(\rho),
%	\label{8.01} \\
\nonumber \\
        f^{-}_g(x,\mu^2) &=& -\frac{4}{9} A_q e^{- d_{-} s} \, + \, O(x),~~
        %\nonumber \\
%	\label{8.00} \\
	f^{-}_q(x,\mu^2) ~=~  A_q e^{-d_{-}(1) s} \, + \, O(x),
	\label{8.02}
\end{eqnarray}
where $I_{\nu}$ ($\nu=0,1$)
%$\tilde{I}_{\nu}$ ($\nu=0,1$)
are the modified Bessel functions
with
\be
\sigma = 2\sqrt{\left|\hat{d}_+\right| s
  \ln \left( \frac{1}{x} \right)}  \, ,~~ \rho=\frac{\sigma}{2\ln(1/x)},~~
  \hat{d}_+ = - \frac{12}{\beta_0},~~
\overline d_{+} = 1 + \frac{20f}{27\beta_0},~~
d_{-} = \frac{16f}{27\beta_0} \, .
\label{intro:1a}
%\eea
\ee
Here the factors $\mu_0^2$, $A_a$ are free parameters obtained in~\cite{Kotikov:2017mhk}
and given there in Table~1.
%(see Table~1 in Ref.~\cite{Kotikov:2017mhk}).

\subsection{Valence part}

For the valence part we have (see, for example, Ref.~\cite{Illarionov:2010gy} and references therein)
\begin{equation}
\tilde{f}_q^V(x,\mu^2) =  A_V(s) \, x^{\lambda_V} \,,~~
A_V(s) = A_V(0) e^{-d_{NS}(1-\lambda_V) s} \,,
% ~~~(i=V,NS),
\label{S3.1}
\end{equation}
where
\be
%A_V(s) = A_V(0) e^{-d_{NS}(1-\lambda_V) s} \,,
%\label{S3.2} \\
d_{NS}(n) =
%\frac{32}{3}
\frac{16}{3\beta_0} \left[
\Psi(n+1)+ \gamma_E -\frac{3}{4}- \frac{1}{2n(n+1)} \right] \,
%\nonumber
\label{S3.2}
\ee
%where
and the quantities $\lambda_V$ and $A_V(0)$ are free parameters obtained in~\cite{Illarionov:2010gy}
(quoted there in Table~1)
%(see Table~1 in~\cite{Illarionov:2010gy})
and $\Psi(n+1)$ is a Euler $\Psi$-function.

It is also interesting to consider parametrization of the valence part proposed in~\cite{Gluck:2007ck},
where it is expressed as a combination of the corresponding contributions of $u$ and $d$ quarks:
\begin{equation}
\tilde{f}_q^V(x,\mu^2) = \sum_{i=1,2} \, \tilde{f}_{q_i}^V(x,\mu^2), ~~
\tilde{f}_{q_i}^V(x,\mu^2) =  A_V^i(s) \, x^{\lambda_V^i}\,, (q_1=u,~q_2=d) \,.
\label{S3.3}
\ee
%where $q_1=u$ and $q_2=d$.
The values for $A_V^i(0) \equiv N_i$
and $\lambda_V^i \equiv a_i$ can be found in Table~1 of~\cite{Gluck:2007ck}.

We note that all the results will be obtained by using the analytic coupling constant
\cite{Shirkov:1997wi} (see discussion in~\cite{Kotikov:2017mhk}), which usually
leads to stable results at low $\mu^2$ values \cite{Kotikov:2012sm}.

Note also that the equations (\ref{8.02}),
%and
(\ref{S3.1}) and (\ref{S3.3}) are in principle dealing with different $s$ values,
because results obtained in~\cite{Illarionov:2010gy}, \cite{Q2evo} and \cite{Gluck:2007ck} contain
similar but not equal values of
the parameters of $\mu^2$-evolutions.
However, it is not as important for the presentation and therefore in what follows we keep always the same parameter $s$.

\section{Rescaling model}

In the rescaling model~\cite{Close:1984zn} SF $F_2(x,\mu^2)$ and
parton densities,
are modified by rescaling $\mu$ variable in the case of a nucleus $A$.
Note that it is usually convenient to study the following ratio
%(see Fig.~1 in Ref.~\cite{Kulagin:2016fzf})
\begin{equation}
R^{AD}_{F2}(x,\mu^2) = \frac{F^A_2(x,\mu^2)}{F^D_2(x,\mu^2)}\,,
\label{AD}
\end{equation}
since the nuclear effect in a deuteron is very small
% (see Table~1 for the values of $\delta^A_{v}$ and discussions in~\cite{Kulagin:2016fzf})
\footnote{The study of nuclear effects in a deuteron can be found in the recent paper~\cite{Alekhin:2017fpf},
  which also contains short reviews of preliminary investigations.}.

We can suggest that
\begin{eqnarray}
  F^D_2(x,\mu^2) &=& e \,
  %f_q(x,\mu^2),~~
 \bigl(
  f_q^S(x,\mu^2)+f_q^V(x,\mu^2) \bigr),~~
  %\nonumber \\
  F^A_2(x,\mu^2) =
  %&=&
  e \,
  %\overline{f}^{A}_q(x,\mu^2),
  \bigl(
  f_q^{AS}(x,\mu^2)+f_q^{AV}(x,\mu^2) \bigr) \, ,
    \nonumber \\
\tilde{f}^{AS}_a(x,\mu^2) &=&
%~=~
f_a^{A,+}(x,\mu^2) + f_a^{A,-}(x,\mu^2),~~
f^{A,\pm}_a(x,\mu^2) =
%&=&
f^{\pm}_a(x,\mu^2_{AD,\pm}) \, ,  \nonumber \\
\tilde{f}^{AV}_q(x,\mu^2) &=& \tilde{f}^{V}_q(x,\mu^2_{AD,v})\,,
\label{AD1}
\end{eqnarray}
where $f^{\pm}_a(x,\mu^2)$ and $\tilde{f}^{V}_q(x,\mu^2)$ are given in Eqs. (\ref{8.02}), (\ref{intro:1a}) and (\ref{S3.1}) with
\be
s^{AD}_{k} \equiv \ln \left(\frac{\ln\left(\mu^2_{AD,k}/\Lambda^2\right)}{\ln\left(\mu^2_{0}/\Lambda^2\right)}\right)
  = s +\ln\Bigl(1+\delta^{AD}_{k}\Bigr),~~(k=\pm,v)
\label{sk}
\ee
i.e. the nuclear modification of the basic variable $s$ depends only on the
%$\mu^2$ independent
parameters $\delta^{AD}_v$ and $\delta^{AD}_{\pm}$, which are $\mu^2$, $\mu_0^2$ and $\Lambda$ independent
(see~\cite{Kotikov:2017mhk}).

The results for $\delta^{AD}_{\pm}$ and $\delta^{AD}_{v}$ are obtained in~\cite{Kotikov:2017mhk}.
In the lead case they are as follows:
\be
\delta^{AD}_{+}= -0.346 ,~~ \delta^{AD}_{-}= -0.779 ,~~ \delta^{AD}_{v}= 0.14 \, .
\label{d.i}
\ee

\section{Results}

\begin{figure}[t]
\centering
\vskip 0.5cm
\includegraphics[height=0.45\textheight,width=0.8\hsize]{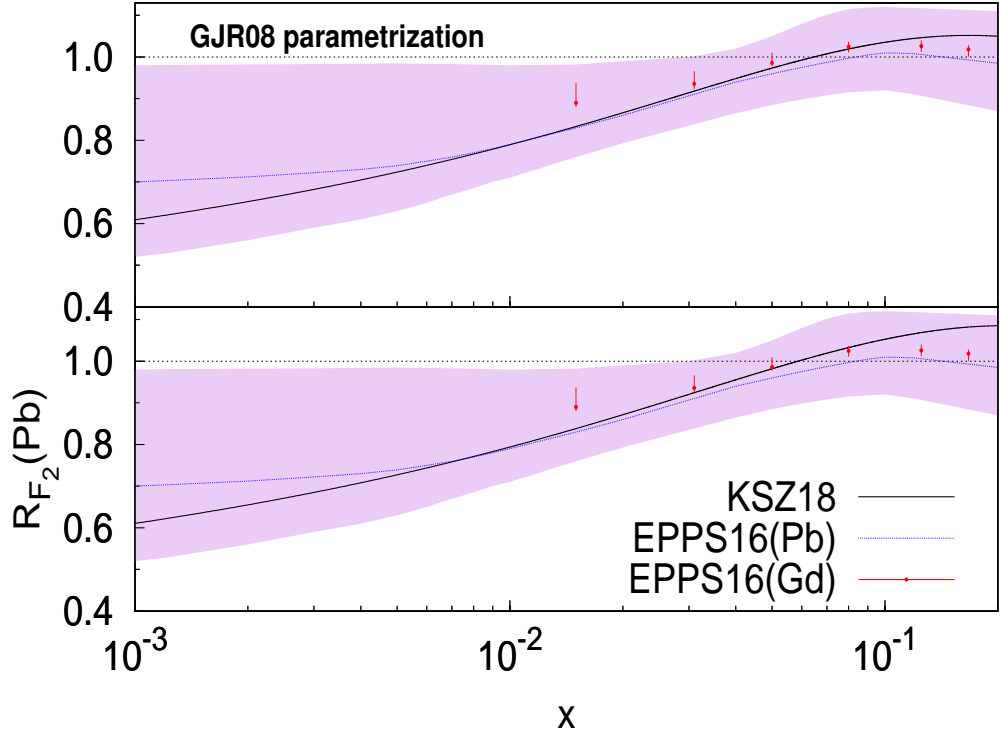}
\vskip -0.3cm
\caption{$x$ dependence of $R^{AD}_{F2}(x,\mu^2)$
%and $R^{AD}_{g}(x,\mu^2)$
at $\mu^2$=10 GeV$^2$ for lead data.
A blue curve with pink band (shows 90$\%$ uncertainties) and red points
(obtained for gadolinium $A=64$ and properly scaled up to fit lead data)
are borrowed from~\cite{Eskola:2016oht} (see Figs. 27 and 13),
%the second paper of~\cite{Armesto:2006ph},
while a black one is obtained in the present paper.}
%\label{fig:F2}
\end{figure}

A positive modification of $R^{AD}_{F2}(x,\mu^2)$ at $x \sim 0.1$ was predicted \cite{Nikolaev:1975vy}
by using the momentum sum rule.
Similar behavior was also suggested \cite{Brodsky:1989qz} on the basis of interference in the multiple scattering description.

The Fermilab E772 Drell-Yan experiment indicated \cite{Alde:1990im} no nuclear modification in antiquark
distributions, so that the antishadowing in $R^{AD}_{F2}$ should be ascribed to the valence quark modifications.
It is demonstrated in Fig.~1 by EPPS16 results~\cite{Eskola:2016oht}, where the blue line shows a sea contribution
while the red points represent a complete contribution.
It is this observation that was behind the motivation to consider the valence density $f_q^V(x,\mu^2)$ in the present analysis.

The obtained results for $R^{AD}_{F2}(x,\mu^2)$, which are for $x \leq 10^{-2}$ very close
to those derived in \cite{Kotikov:2017mhk},
 show (see Fig. 1)
 an appearance of the antishadowing effect for $x \sim 0.06$ and its rise with increasing
 $x$ values all the way up to $x \geq 0.1$, which is a limit of the current consideration.
 The results are actually shown up to $x \sim 0.2$; however, for the values higher than $x \sim 0.1$ they cannot be as accurate.

The results obtained for two differently parametrized valence quark densities,
proposed in~\cite{Illarionov:2010gy} and~\cite{Gluck:2007ck}, are close to each other.
It is seen that GJR08 parametrization exhibits a weaker antishadowing effect. Both curves lie a bit higher than the EPPS16 ones.
%%% (see red points in Fig.~1, which were obtained for gadolinium data, that should little decrease of the effect to compare with lead).
However, all the results are completely consistent within uncertainties of the EPPS16 analysis (see the pink band).

%\begin{acknowledgments}
Support by the National Natural Science Foundation of China (Grant
No. 11575254) is acknowledged. AVK and BGS thank Institute of Modern Physics
for invitation. AVK is also grateful to the CAS President's International Fellowship Initiative
(Grant No.~2017VMA0040) for support.
The work of AVK and BGS was in part supported by the RFBR Foundation through the Grant No.~16-02-00790-a.
%\end{acknowledgments}

\end{document}